\documentclass[12pt]{article}
\pdfoutput=1
\usepackage{graphicx}
\usepackage{amsmath}
\usepackage{makeidx}

\usepackage{a4,color,graphics}
\usepackage{amsfonts}
\usepackage{amssymb}
\usepackage{graphicx}%
\usepackage[all]{xy} 

\makeatletter \@addtoreset{equation}{section}

\def\be{\begin{equation}}
\def\ee{\end{equation}}
\def\bea{\begin{eqnarray}}
\def\eea{\end{eqnarray}}

\newcommand{\nc}{\newcommand}
\nc{\al}{\alpha} \nc{\bib}{\bibitem} \nc{\la}{\lambda}
\nc{\C}{\mbox{\hspace{1.24mm}\rule{0.2mm}{2.5mm}\hspace{-2.7mm}
C}} \nc{\R}{\mbox{\hspace{.04mm}\rule{0.2mm}{2.8mm}\hspace{-1.5mm}
R}}

\begin{document} 
\title{%
\textbf{ Bilocal theory and gravity I }} 
\author{ Pablo Diaz\thanks{pablo.diazbenito@uleth.ca}, ~Saurya Das\thanks{saurya.das@uleth.ca} ~and Mark Walton\thanks{walton@uleth.ca} \\
{\small \emph{Theoretical Physics Group,
Department of Physics and Astronomy, University of Lethbridge,}}\\
{\small \emph{4401 University Drive, Lethbridge, Alberta, T1K 3M4, Canada}}
}

\maketitle
 
\begin{abstract}
With the aim of investigating the relation between gravity and non-locality at the classical level, we study a bilocal scalar field model. Bilocality introduces new (internal) degrees of freedom which reproduces gravity in the following sense: 
we show that the equations of motion of the 
massless branch of the free bilocal model match those of linearized gravity in a specific gauge,
and so does their solutions. 
We also discuss higher orders in perturbation theory, 
where there is self-interaction in both gravity and the bilocal field sectors.\\

\textbf{Keywords}: Gravity, bilocal field, non-locality, perturbative gravity, gravitational waves.
\end{abstract}
\thispagestyle{empty} 
\newpage
\pagestyle{plain} 

\tableofcontents
\newpage


\section{Introduction}

\label{sec: introduction}

It is believed that resolving two spacetime points is impossible when they are sufficiently close to each other. 
A simple argument goes as follows: resolving two nearby points amounts 
to probing that region of spacetime with particles of wavelength of the order of the distance we want to resolve. As we consider closer points the wavelength must be shorter, with
more and more energetic probes. This process cannot go on forever. According to general relativity, if enough radiation is aimed into a region, the concentration of energy warps spacetime and the region becomes a black hole. The event horizon of the black hole prevents us from resolving any points beyond it \cite{maggiore}.

The theoretical impossibility of resolving arbitrarily  small distances may indicate a fundamental non-locality, as also pointed out in \cite{DFE,Ah}.  
In other words, semiclassical gravity suggests non-locality at a fundamental level. 
This motivates us to ask the reverse question: 
does non-locality imply, or accommodate, gravity in a natural way? 
An affirmative answer to this question would provide a novel, non-local framework in which general relativity is an effective, emergent, low-energy theory. 
Note that the effective nature of gravity has been conjectured long ago 
\cite{emergent}.

We investigate the possibility of gravity being induced by non-locality in the current article. 
We provide significant evidence that non-locality naturally gives rise to gravity at the classical level. The basic requirements
we impose to a non-local theory is that its non-locality must be genuine and short-ranged. 
Genuine non-locality means that the field cannot be decomposed into local quantities. This is implemented by imposing an action at a distance. Short-range non-locality means that the non-local effects must vanish at low energies, where we expect to recover local effective theories. This is a physically reasonable assumption, since local theories are tested with a high accuracy at lab energies, at least up to about $10$ TeV or equivalently $10^{-19}$ m. 

In order to support the claim that non-locality implies gravity we study the simplest type of non-local theories and we compare it with pure gravity (The coupling with matter should not present extra difficulties and will be considered elsewhere). Specifically, we consider {\it bilocal}  classical scalar field theories. Bilocality is the minimum departure from locality in which the fields depend on two spacetime points instead of one. Other more involved non-local theories are expected to lead to effective bilocal theories when suitable degrees of freedom  are integrated out. Therefore the analysis of this paper applies not merely to bilocal theories but to any non-local theory which may lead to an effective bilocality. 

As a result, in this article we study the minimal free model for a scalar field that provides genuine and short-range bilocality.  We see that bilocality opens up new (internal) degrees of freedom that match exactly to those of gravity, at least to first order in perturbation theory. We show explicitly how general massless solutions of the bilocal model match one-to-one with gravitational waves. Thus, the bilocal model is proven to reproduce linearized gravity in the transverse-traceless gauge. This is a highly non-trivial (and the most important) result of the paper, shown in equations (\ref{massless}) and (\ref{goodsols}). The connection between the bilocal theory and linearized gravity works surprisingly well. For instance, in \ref{AS}, we show how the bilocal field solutions (which do not have helicity) develop a $\pi/4$-pattern, identical to the ``plus'' and ``cross'' polarization of gravitational waves, by simply insisting on keeping the bilocal solutions short-ranged. The close relation between linearized gravity and the free bilocal field suggests that a bilocal field theory with a suitable potential could successfully describe full gravity.  We  sketch how this match should hold in subsequent orders in perturbation theory, where there is self-interaction in both gravity and in the bilocal field. 

Bilocal fields were first studied by Yukawa in a series of papers \cite{Yu}. The physical motivation was to describe mesonic excitations. The two spacetime points the field depends on were the location of the quarks. The use of bilocal and trilocal field models to explain confinement in hadrons became popular in the next couple of decades \cite{T,bilocal,Fe}, until the success of quantum chromodynamics made them fade away. In the context of the  AdS/CFT correspondence (in the higher-spin version), bilocal fields have also been reconsidered by Jevicki {\it et al.} during the last decade \cite{jevicki,dMJJR}. 

This paper is organized as follows. Section \ref{LG} is a brief review of linearized gravity. In section \ref{BFM} we present and discuss the free bilocal field model. 
Subsections \ref{sec:massless} and \ref{AS} describe its massless solutions, which constitute 
the main results of this paper. 
Section \ref{GFBF} is devoted to the precise matching between the solutions of both theories; from the similarities of the space of solutions it is easy to infer how quantities of both theories should relate to each other. Such relation is shown in \ref{FO}.
The associations that naturally arise at first order in perturbation theory should hold at higher orders. In \ref{SHOPT} we sketch the way in which both theories should be compared perturbatively. Basically, the inclusion of self-interacting terms in the gravity side should generalize the free bilocal model to the one described by equations (\ref{genmodel}) and (\ref{genconstraint}), for suitable potentials. Before the concluding section we 
also present a short section \ref{PIBF}, where we speculate about the physical meaning of the  bilocal field. 

\section{Linearized gravity}
\label{LG}

First order equations of Einstein gravity are well-known and lead to the gravitational wave equations in the appropriate gauge, see for instance section 4.4 of \cite{RW}. Linearized gravity equations are of the form
\begin{equation}\label{lineargen}
-\Box h_{\mu\nu}+h^{\phantom{1}\alpha}_{\nu \phantom{1},\mu\alpha}+h^{\phantom{1}\alpha}_{\mu \phantom{1},\nu\alpha}-h_{,\mu\nu}-h^{\alpha\beta}_{\phantom{1},\alpha\beta}\eta_{\mu\nu}+\eta_{\mu\nu}\Box h=16\pi G T_{\mu\nu}.
\end{equation}
In this work we will consider only pure gravity, so $T_{\mu\nu}=0$. Upon taking {\it trace-reversed} variables
\begin{equation*}
\bar{h}_{\mu\nu}=h_{\mu\nu}-\frac{1}{2}h\eta_{\mu\nu},
\end{equation*}
equation (\ref{lineargen}) reads:
\begin{equation}\label{tracereversedeq}
-\Box \bar{h}_{\mu\nu}+\bar{h}^{\phantom{1}\alpha}_{\nu \phantom{1},\mu\alpha}+\bar{h}^{\phantom{1}\alpha}_{\mu \phantom{1},\nu\alpha}-\bar{h}^{\alpha\beta}_{\phantom{1},\alpha\beta}\eta_{\mu\nu}=0.
\end{equation}
Taking the Lorentz gauge condition
\begin{equation}\label{lorentz}
\partial^{\mu}\bar{h}_{\mu\nu}=0,
\end{equation}
reduces (\ref{tracereversedeq}) to the simple wave equation
\begin{equation}\label{box}
\Box \bar{h}_{\mu\nu}=0.
\end{equation}
Equation (\ref{lorentz}) does not fix completely the gauge.  The reason is that we can always gauge transform $h_{\mu\nu}\to h_{\mu\nu}+\xi_{\mu,\nu}+\xi_{\nu,\mu}$ with $\Box\xi_{\mu}=0$ and (\ref{lorentz}) still holds \cite{RW}.  \\ 
Imposing the transverse-traceless gauge condition it can be shown that the general solution of (\ref{box}) can be written as
\begin{equation}\label{hbarlinear}
\bar{h}_{\mu\nu}=\Re \big(B_{\mu\nu}e^{iPx}\big),
\end{equation}
where $P_\mu$ is a null vector,  and
\begin{equation}\label{Alinear}
\mathbf{B} = \left(
\begin{array}{cccc}
0 & 0 & 0&0 \\
0& B_+ & B_{\times}&0 \\
0& B_{\times}&-B_+&0 \\
0&0&0&0
\end{array} \right).
\end{equation}
The two remaining degrees of freedom, i.e. the two polarizations, are thus associated to $B_+$ and $B_{\times}$. They are physical degrees of freedom since the gauge is completely fixed.

\section{Bilocal field model} \label{BFM}

Bilocal field models were first proposed by Yukawa \cite{Yu} in the early 50's to describe mesons and confinement. They were extensively used, with slight modifications, in the context of the strong interaction during the two following decades, see \cite{T,bilocal,Fe} and the reference therein. The two spacetime points of the bilocal field were the positions of the quarks in a meson. Mesons, as massive particles, corresponded to massive solutions of the several bilocal models they used. 

The bilocal model we consider in this article is similar to those found in the above references except for two main differences. Our model is minimal, in the sense that it consists of the minimum number of terms and derivatives which are necessary to retain the short-range genuine bilocality. \\
Besides, our model can accommodate massless solutions, crucial for our analysis, since gravitons are massless.     
As far as we know, massless solutions were not considered in the earlier papers. \\

The basic object of a bilocal model is the bilocal field
\begin{equation*}
\bar{\Phi}(x,y),
\end{equation*} 
which as indicated, depends on two points of spacetime as opposed to a single point as in local theories. It is common to work with a different set of 
coordinates which are more physical. The {\it centre of mass} (CM) coordinates,
\begin{equation*}\label{coor}
X_{\mu}=\frac{1}{2}(x_{\mu}+y_{\mu}), 
\end{equation*}
will be the ones that can be observed, i.e. identifiable with the regular macroscopic
coordinates, and survive in the local limit.
The  {\it relative} (or {\it internal}) coordinates on the other hand 
are related to the distance between $x$ and $y$. 
We will define the ``Euclidean'' version of this distance. 
This avoids physical inconsistencies 
associated with having `two times', as well as the violation of causality,
which would accompany action-at-a-distance in Lorentzian internal coordinates. 
\\

%
\begin{equation}\label{relativecoor}
 r_{\mu}=(x_{\mu}-y_{\mu})_E,\quad r^2=r^{\mu}r^{\nu}\eta_{\mu\nu}\geq 0.
\end{equation}
%
%
The field in the new coordinates will be called ${\Phi}(X,r)$, with the definition
\begin{equation*}
\Phi(X,r)=\bar{\Phi}(x,y).
\end{equation*}
We will work in four dimensions, and assume $\mu=0,1,2,3$. As is customary, we consider the bilocal field symmetric under the exchange $x\leftrightarrow y$. That is, we consider $\bar{\Phi}(x,y)=\bar{\Phi}(y,x)$ or, equivalently, 
\begin{equation}\label{evencond}
\Phi(X,r)=\Phi(X,-r).
\end{equation}
%
%
This ensures that neither of the internal points $x$ or $y$ is preferred, and that
physics is invariant under their interchange.
We see that the effect of the bilocal field is to bring in
new degrees of freedom, namely the dependence of the field on the relative coordinates. We will show that it is precisely in those degrees of freedom that gravity is encoded.

The classical bilocal scalar field model we are considering consists of two equations:
%
\begin{eqnarray}
\big(\Box+\Box_r-\alpha^4r^2+2\alpha^2\big)\Phi(X,r)&=&0 \label{eom}\\
\frac{\partial}{\partial X_{\mu}}\bigg(\frac{\partial}{\partial r^{\mu}}+\alpha^2 r_{\mu}\bigg)\Phi(X,r)&=&0 \label{constraint},
\end{eqnarray}
where $\Box$ and $\Box_r$ are the d'Alembertian operators 
associated with the coordinates $X_{\mu}$ and $r_{\mu}$ respectively. 
The parameter $\alpha$ has dimensions of mass.  

Equation (\ref{eom}) is dynamical whereas equation (\ref{constraint}) is a constraint, as can be guessed at first sight since eq. (\ref{constraint}) only involves  first derivatives of the CM coordinates. 
%
We can see from the equations (\ref{eom}) and (\ref{constraint}) that the model has the minimal ingredients to be short-range, yet genuinely bilocal.
The term  $\alpha^4r^2$ in Eq.(\ref{eom}) forces short-ranged non-locality, since it makes the general solutions of the model have a Gaussian decay factor in the relative coordinates. 
 

\begin{figure}
\centering
  \includegraphics[scale=0.37]{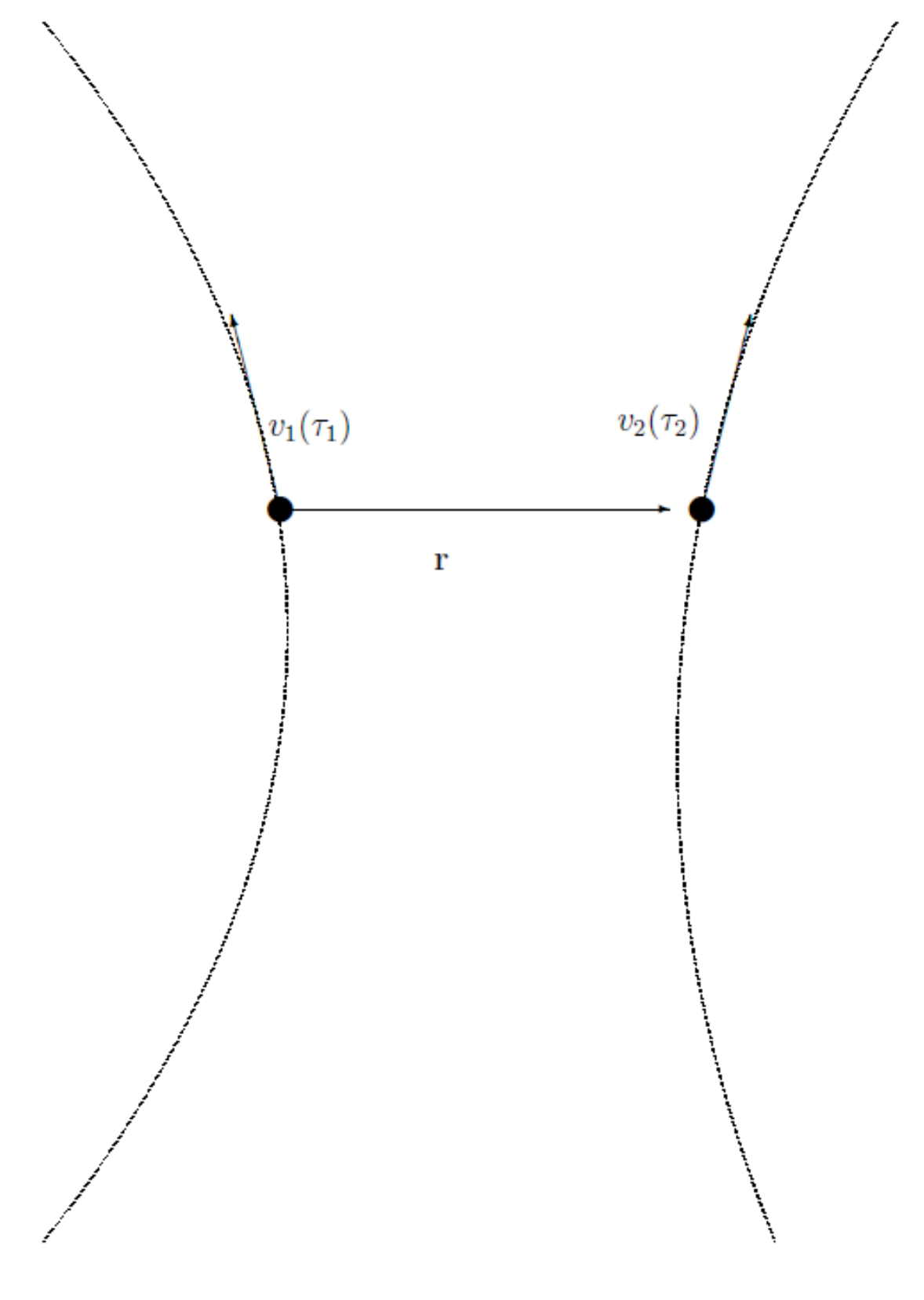}
  \caption{In the point particle picture, the bilocal equations describes the motion of a bound system of two particles. The equations (\ref{eom}) and (\ref{constraint}) translate into a harmonic instant interaction between the two particles.
}
  \label{pointparticle}
\end{figure}

It is easier to understand the implications of these two equations if we shift
for a moment from the field theory to a point-particle description (Fig.\ref{pointparticle}). In this figure, we see the worldlines of the two-particle system that the model describes. Parameters $\tau_1$ and $\tau_2$ are their respective proper times. Now, equation (\ref{constraint}) is derived from the simple kinematical relation ${\bf r}\cdot({\bf v}(\tau_1)+{\bf v}(\tau_2))=0$, which is independent of the reparametrizations of $\tau_1$ and $\tau_2$. 
This geometrical constraint implies a condition $\tau_1(\tau_2)$, 
which in turn implies an action-at-a-distance between the two particles \cite{T2}. In summary, Eq.(\ref{constraint}) serves to identify the proper time corresponding
to the points $x$ and $y$, which in turn ensures an action-at-a-distance.
Note that different constraints (and so different actions-at-a-distance) could have been chosen \footnote{See \cite{T2}, for a comprehensive analysis of these constraints. When comparing with gravity, the constraint equation translates into the Lorentz gauge. It seems reasonable to associate the arbitrariness of the choice of constraint with the freedom of choosing different gauges in gravity.}. 
Eq.(\ref{eom}) and Fig.\ref{pointparticle} show
that the particles interact as if linked by a spring, with the spring
constant controlled by parameter $\alpha$.   \\

It is generally accepted that gravitons are massless. Thus, we will study the general massless solutions of the model (\ref{eom}) and (\ref{constraint}) in the next section, and we will show how they naturally encode gravitational waves at linear order. \\
%


\subsection{Massless solutions}
\label{sec:massless}

For solutions of (\ref{eom}) and (\ref{constraint}) we take the ansatz 
\begin{equation}
\label{massless}
\Phi_{\lambda,P}(X,r)=e^{iPX}\text{exp}\Big[-\frac{\alpha^2}{2}A_{\mu\nu}r^{\mu}r^{\nu}\Big]f_{\lambda}( {\bf r}),
\end{equation}
where $P^{\mu}$ is a null vector, which indicates that the solutions are massless. As customary \cite{T,bilocal,Fe}, the mass attribute of the field is related to the CM coordinates. In our case, the bilocal system, as a whole, behaves as a massless particle. \\
This vector can be written as $P=p(1,\vec n)$,
where  $\vec{n}$ is a three-dimensional spatial vector such that $\delta_{ij}n^in^j=1$. Without loss of generality, we will choose from now on the unit vector $\vec{n}$ parallel to the $z$-axis, so $P=(p,0,0,p)$.
For later use, we define the null vector $K=(-1/p,0,0,1/p)$ propagating along the negative $z$-axis. The index $\lambda$ (or a collection thereof) labels 
polynomials of the relative coordinates.  

From the ansatz (\ref{massless}) we see that the parameter $\alpha$  measures non-locality. The limit
\begin{equation*}
\lim_{\alpha\to \infty}\frac{\alpha}{\sqrt {2\pi}}e^{-\frac{1}{2}\alpha^2 x^2}=\delta(x),
\end{equation*} 
applied to each coordinate, implies that the relative dimensions vanish for large $\alpha$. So,  in the limit $\alpha\to \infty$ the space of relative coordinates shrinks and one is left with a local theory.\\
%
We search for solutions of (\ref{eom}) and (\ref{constraint}) of the form (\ref{massless}).  Although $\Phi$ is a scalar, spin-0 field, its non-locality leads to the Gaussian factor in the internal coordinates in (\ref{massless}). This factor requires the existence of a symmetric tensor $A_{\mu\nu}$, contracted with the relative coordinates, whose general shape will be given below. \\
%
Thus an internal structure of the scalar field emerges due to bilocality, whose
dynamics will match with those of linearized gravity, as
we will show in the following sections. \\

First note that, since $P_{\mu}$ is a null vector, 
ansatz (\ref{massless}) satisfies 
\begin{equation*}
\Box \Phi_{\lambda,P}(X,r)=0.
\end{equation*}
Therefore, it follows from Eq.(\ref{eom}), that the following must hold:
\begin{equation}\label{remains}
\big(\Box_r-\alpha^4r^2+2\alpha^2\big)\Phi_{\lambda,P}(X,r)=0,
\end{equation}
which in turn implies:
\begin{eqnarray}
\label{Rproduct}
A_{\mu\sigma}A^{\sigma}_{\nu}&=&\eta_{\mu\nu}, \\
\eta^{\mu\nu}A_{\mu\nu}&=&2.
\label{Rproduct2}
\end{eqnarray}
Next from Eq.(\ref{constraint}) one gets
\begin{equation}
P^{\mu}A_{\mu\nu}=P_{\nu}.
\label{PA}
\end{equation}
An ansatz which implements conditions (\ref{Rproduct}), (\ref{Rproduct2}) 
and (\ref{PA}) is
%
\begin{equation}
\label{goodsols}
{\bf A}=\left(
\begin{array}{cccc}
-1& 0 & 0 &0\\
0&a & b&0\\
0&b&-a&0\\
0& 0 & 0 &1
\end{array} \right),\quad  \det {\bf A}=1,
\end{equation}
with $a,b$  real numbers. We will show later that 
Matrix ${\bf A}$ will be naturally associated with gravitational waves.
Eqs.(\ref{massless}) and (\ref{goodsols}) are two important results of the paper, and
it follows that these general massless solutions of the simplest non-local free scalar model are essentially gravitational waves.  

Plugging (\ref{massless}) into (\ref{eom}) we see that the linear equation for functions $f_{\lambda}$ is 
%
\begin{equation}\label{feq}
\big(-2\alpha^2r^{\sigma}A_{\sigma\mu}\partial^{\mu}+\Box_r\big)f_{\lambda}( r)=0.
\end{equation}
We see that $f_\lambda=1$ solves the equation. 
In general, $f_\lambda$  are nontrivial polynomials involving the parameters $a$ and $b$.  
We leave the study of the general solutions of (\ref{feq}) to a future publication, 
but an example of such polynomials is
\begin{equation*}
f(r)=\alpha^2(-br_1^2+br_2^2+2ar_1r_2).
\end{equation*}

Next we examine short range solutions, which must vanish when $r \to \infty$, which requires 
the matrix ${\bf A}$ to be positive definite:
\begin{equation}\label{posdef}
r^{\mu}r^{\nu}A_{\mu\nu}> 0.
\end{equation} 
Solutions (\ref{massless}) with (\ref{goodsols}) 
are not short-range as they stand. Using (\ref{goodsols}) we see that 
\begin{equation}\label{signs}
r^{\mu}r^{\nu}A_{\mu\nu}=-r_0^2+r_3^2+ar_1^2-ar_2^2+2br_1r_2.
\end{equation} 
Coordinates $r_0$ and $r_3$ are not problematic, since $-r_0^2+r_3^2$ is always positive
\footnote{Recall that $r_0=i(x_0-y_0)$ as defined in (\ref{relativecoor}).}
%
The $(r_,r_2)$-plane is more subtle, since for example Eq.(\ref{signs}) causes 
(\ref{massless}) to blow up for $a>0$ and $r_2\to\infty$. 
However, as we will show in the next subsection, global short-ranged solutions can be constructed
based on (\ref{goodsols}). We will see that these solutions develop a natural 
$\pi/4$-pattern which mimics the `plus' and `cross' polarizations of gravitational waves.

\subsection{Short-ranged solutions}
\label{AS}

In this section we construct short-range solutions of the form (\ref{massless}) defined on the whole $(r_1,r_2)$-plane. We will call them global solutions. Since the following
discussion  holds for a general $f_{\lambda}(r)$, for simplicity we take $f_{\lambda}=1$. We write $\Phi_{(a,b)}(X,r)$ for the solution 
with $A_{\mu\nu}$ as in (\ref{goodsols}).

As mentioned above, 
$r^{\mu}r^{\nu}A_{\mu\nu}>0$ guarantees that the functions decay as Gaussians. 
Without loss of generality we take $a,b\geq 0$ and $r_0=r_3=0$ in this section. 
Analyzing the sign of $r^{\mu}r^{\nu}A_{\mu\nu}$ in (\ref{signs}) we see that 
\begin{equation}
r^{\mu}r^{\nu}A_{\mu\nu}> 0 \longrightarrow r_1> cr_2,\quad c=\frac{-b}{a}+\sqrt{1+\frac{b^2}{a^2}}.
\end{equation}
The constant $c$ takes values in the interval $(0,1)$, so for the wedge $r_1\geq r_2\geq 0$ the sign of $r^{\mu}r^{\nu}A_{\mu\nu}$ is always positive, and the solutions $\Phi_{(a,b)}(X,r)$ are short-range in this wedge. Similarly, the function $\Phi_{(-a,b)}(X,r)$ makes the product $r^{\mu}r^{\nu}A_{\mu\nu}$ always positive in the wedge $r_2\geq r_1\geq 0$. 
Therefore one is tempted to consider the solution $\Phi_{(a,b)}(X,r)$ if $r_1\geq r_2\geq 0$, and as $\Phi_{(-a,b)}(X,r)$ if $r_2\geq r_1\geq 0$, so that
it behaves well at infinity in the entire first quadrant. 
However we have to make sure that solutions are continuous and with continuous derivatives all over the plane, and the above choice would not have continuous derivatives on the line $r_1=r_2$, as one can check in (\ref{signs}). 
Therefore we impose continuity of the solutions and of the directional derivative $\partial_{r_2}-\partial_{r_1}$ of the solutions along the line $r_1=r_2$. 
It is easy to see that the solution
\begin{eqnarray*}
&\Phi_{(a,b)}(X,r)-\Phi_{(0,b)}(X,r),&\quad r_1> r_2\geq 0 \nonumber \\
&-\Phi_{(-a,b)}(X,r)+\Phi_{(0,b)}(X,r),&\quad r_2\geq r_1\geq 0
\end{eqnarray*}
is continuous and has continuous derivatives everywhere, 
so it is a well behaved function defined on the first quadrant. 
Next, we notice that functions $\Phi_{(-a,-b)}(X,r)$ decay as Gaussians in the wedge $|r_1|<r_2$ and $r_1\leq 0$, and again we can find the way of gluing it together properly along the line $r_1=0$. This can be done till we complete the circle and the solutions are defined on the entire $(r_1,r_2)$-plane.
So, for any $a,b\geq 0$ we find a global solution
\begin{equation}\label{global2}
\Phi^{G}_{(a,b)}(X,r) = \left\{
\begin{array}{ll}
\Phi^{I}_{(a,b)}\equiv\Phi_{(a,b)},& |r_1| \geq |r_2| \text{ and } r_1\cdot r_2 > 0,\\
\Phi^{II}_{(a,b)}\equiv -\Phi_{(-a,b)}+2\Phi_{(0,b)},& |r_2| > |r_1| \text{ and } r_1\cdot r_2\geq 0,\\
\Phi^{III}_{(a,b)}\equiv \Phi_{(-a,-b)}-2\Phi_{(0,-b)}-2\Phi_{(-a,0)}+4, & |r_2| \geq |r_1| \text{ and } r_1\cdot r_2< 0,\\
\Phi^{IV}_{(a,b)}\equiv -\Phi_{(a,-b)}+2\Phi_{(a,0)}, &  |r_1| > |r_2| \text{ and } r_1\cdot r_2\leq 0.
\end{array} \right.
\end{equation} 
Solutions (\ref{global2}) are shown in the figure and are everywhere regular.\\
It can also be shown that the global solutions (\ref{global2}) are essentially the only choice to assemble a short-range solution. Note that, as said above, this discussion holds for any function $f_{\lambda}( r)$ since the Gaussian behaviour dominates over any polynomial for large values of $r$. Because our free model is linear, linear superpositions lead to other global solutions. So, a general global solution can be
expressed as a linear combination of the above solutions and, then, written as 
\begin{equation}\label{SG}
\Phi^G_A(X,r)=\int_{a,b\in \Re_+} A(a,b)\Phi^G_{(a,b)}(X,r)~da~ db,
\end{equation}
%
for any real square-integrable function $A(a,b)$.\\
\vspace{1.cm}
\setlength{\unitlength}{1.7cm}
\begin{picture}(10,4)(-4,-2.1)
\put(-2.4,0){\vector(1,0){4.9}}
\put(2.7,-0.1){$r_1$}
\put(2.7,-0.1){$r_1$}
\put(0,-1.9){\vector(0,1){3.9}}
\put(-1.9,-1.9){\line(1,1){3.8}} 
\put(1.85,1.6)
{$r_1=r_2$}
\put(1.85,1.6)
{$r_1=r_2$}
\put(-1.9,1.9){\line(1,-1){3.8}}
\put(1.85,-1.65)
{$r_1=-r_2$}
\put(1.85,-1.65)
{$r_1=-r_2$}
\put(0.1,1.8)
{$r_2$}
\put(0.1,1.8)
{$r_2$}
\put(1.2,0.6) {$\Phi_{(a,b)}^{I}$}
\put(-1.8,-0.6){ $\Phi_{(a,b)}^{I}$}
\put(0.2,1.4){$\Phi_{(a,b)}^{II}$}
\put(-1,-1.4){ $\Phi_{(a,b)}^{II}$}
\put(-1,1.4){ $\Phi_{(a,b)}^{III}$}
\put(0.2,-1.4){ $\Phi_{(a,b)}^{III}$}
\put(-1.8,0.6){ $\Phi_{(a,b)}^{IV}$}
\put(1.2,-0.6){ $\Phi_{(a,b)}^{IV}$}
\put(-3.8,-2.3){{\small 
\text{{\small The global solution as we glue the 8 wedges. Note the symmetry $\Phi^G_{(a,b)}(X,r)=\Phi^G_{(a,b)}(X,-r)$.} }}}
\end{picture}
%
Note that how these solutions, when forced to be short-ranged, develop a $\pi/4$-pattern reminiscent of gravitational waves. 
Whereas in the bilocal model there is no natural concept of helicity, 
the $\pi/4$ pattern of the global solution is forced by the requirement of
short-ranged non-locality. 
Then the internal space `mimics' a spin-2 field this way, and it reproduces the well-known $\pi/4$ pattern of the ``cross'' and ``plus'' polarizations of gravitational waves.

\section{Gravity from a bilocal scalar field}
\label{GFBF}

At this stage, as the reader may guess, some identifications need to be 
established between the bilocal theory and linearized gravity. 
For instance, it is clear that matrix ${\bf A}$ in EQ.(\ref{goodsols}) 
must be related to matrix ${\bf B}$ in (\ref{Alinear}). 
The purpose of this section is to establish a formal matching between both theories.
This will help us understand in detail how free bilocal model encodes linearized gravity and provide clues about how to compare both theories at higher orders in perturbation theory.
 
Let us write our bilocal field in the center of mass 
and relative coordinates $(X,r)$ given by (\ref{coor}) and Taylor expand around
the coordinate $r$ around $r=0$. 
We remind that $\Phi(X,r)$ is even in the relative coordinates. We have 
\begin{equation}\label{Taylor2}
\Phi(X,r)=\phi(X)+H_{\mu\nu}r^{\mu}r^{\nu}+D_{\mu\nu\sigma\rho}r^{\mu}r^{\nu}r^{\sigma}r^{\rho}\cdots,
\end{equation}     
with the identification 
\begin{eqnarray}\label{fieldtower}
\phi(X)&\equiv&\Phi(X,0)  \\
H_{\mu\nu}(X)&\equiv&\frac{1}{2}\partial_{\mu}\partial_{\nu}\Phi(X,r)\Big|_{r=0}  \\
D_{\mu\nu\sigma\rho}(X)&\equiv&\frac{1}{4!}\partial_{\mu}\partial_{\nu}\partial_{\sigma}\partial_{\rho}\Phi(X,r)\Big|_{r=0}  
\\
&\vdots& \label{dots}
\end{eqnarray}
where the partial derivatives are with respect to the coordinate $r$.
This is the way of seeing that the bilocal scalar field is equivalent to a unique tower of {\it local} higher spin fields. We will ignore the role of the higher spin fields in this work and focus on the lowest contribution $H_{\mu\nu}$ of the expansion (\ref{Taylor2}). The field $H_{\mu\nu}$ will make contact with gravity, after imposing the dynamics of the bilocal model (\ref{eom}) and (\ref{constraint}). 

\subsection{First order}
\label{FO}

Equations (\ref{eom}) and (\ref{constraint}) will be considered first order in perturbation theory, with $\kappa_B$ as the perturbation parameter. 
Inserting (\ref{Taylor2}) into either (\ref{eom}) or (\ref{constraint}) produces an infinite tower of equations as we equate powers of $r$. Those equations involve all the higher spin fields although, as mentioned above, we will focus on $H_{\mu\nu}$ in this article. The lowest power of $r$ in (\ref{constraint}) leads to
\begin{equation}\label{Lorentz}
\frac{\partial}{\partial X^\mu}(H_{\mu\nu}+\alpha^2\phi~\eta_{\mu\nu})r^\nu=0 \quad \forall r \longrightarrow \frac{\partial}{\partial X^\mu}(H_{\mu\nu}+\alpha^2\phi~\eta_{\mu\nu})=0,
\end{equation}
where we have used the ansatz (\ref{massless}). Note that since $\Phi\sim e^{iPX}$, constraint (\ref{Lorentz}) only affects the space spanned by $r_0$ and $r_3$, and we get
\begin{equation}\label{Lorentz2}
\frac{\partial}{\partial X^\mu}\big(H_{\mu\nu}+\alpha^2\phi~(PK)_{\mu\nu}\big)=0,
\end{equation}
where
\begin{equation}
(PK)_{\mu\nu}\equiv\frac{1}{2}(P_{\mu}K_{\nu}+P_{\nu}K_{\mu}),
\end{equation}
%
%
with $P$ and $K$ as defined before
\footnote{Note that in the coordinates we are working, the matrix $PK$ takes the simple form {\tiny $\left(
\begin{array}{cccc}
-1& 0 & 0 &0\\
0&0 & 0&0\\
0&0&0&0\\
0& 0 & 0 &1
\end{array} \right)$.}}.
Now, since (\ref{constraint}) is a constraint, it is natural to associate it with a constraint in gravity. With the identification
\begin{equation}\label{hbar}
\frac{-1}{\alpha^2}\Re\big(H_{\mu\nu}+\alpha^2\phi~(PK)_{\mu\nu}\big)\equiv \bar{h}_{\mu\nu},
\end{equation}
%
%
we recover in (\ref{Lorentz2}) the Lorentz gauge condition (\ref{lorentz}) of linear gravity. 
We see that the constraint of the bilocal model corresponds to a (partial) gauge fixing in the linearized gravity side. We also  realize that the center of mass coordinates $X$ of the bilocal field are to be associated with ordinary spacetime coordinates in gravity. 

It is not obvious {\it a priori} that the identification (\ref{hbar}) makes the bilocal dynamics reproduce those of gravity. However, with the definition (\ref{hbar}) and the bilocal solutions (\ref{massless}), the polarization tensor of gravitational waves $B_{\mu\nu}$ in  (\ref{Alinear}) is reproduced in the bilocal side if we identify
\begin{equation}
B_+\equiv \frac{a}{\sqrt{a^2+b^2}}\text{ and } B_{\times}\equiv\frac{b}{\sqrt{a^2+b^2}}
\end{equation}
%
%
%

Here we see that the 
solutions of linearized gravity are encoded in the bilocal field. We would like to stress that the crucial element for this match is the shape of solutions (\ref{goodsols}) of the bilocal field, from which the association (\ref{hbar}) follows naturally.
This is a strong result in our opinion, as the model was chosen for being minimal.

Now we would like to learn about what the gauge in gravity translates into in the bilocal setup. First, let us notice that there is no gauge freedom in our bilocal model. An easy way to see this is to realize that if $A_{\mu\nu}$ leads to a solution of the model, then $A_{\mu\nu}+\partial_{(\mu}\xi_{\nu)}$ does not. This is consistent with the statement that the match with gravity happens in a specific gauge, which is precisely the gauge where gravitational waves are found. We have already seen that the constraint of the bilocal model corresponds to the Lorentz gauge condition, which partially fixes the gauge in gravity. Now, because $\Box\Phi(X,r)=0$ for all bilocal solutions, the rest of the ``gauge fixing''  (the transverse-traceless gauge condition of gravity) must be encoded in the equation (\ref{remains}). Thus, the dynamics of the relative coordinates in the bilocal field model are seen as gauge constraints from the gravity point of view. 

As a consistency check on the gravity gauge that the bilocal model selects, let us note that massless particles must be invariant under the little group $E(2)$ \cite{Wigner}, the subgroup of the Poincar\'e group that stabilizes a null 4-momentum. The group $E(2)$ has three generators, which can be visualized by their actions 
on the 2-plane: one of them generates rotations, and the other two generate translations on the plane. It is known that out of the three generators of $E(2)$ the two associated with translations produce gauge transformations \cite{Kim,JJ} for the electromagnetic field and gravity. The only one which does not affect the gauge is the generator of rotations. Since there is no gauge symmetry in the bilocal model the solutions can only involve $SO(2)\subset E(2)$, as seen from the gravity point of view. The other two generators must act trivially, but this is what actually happens in the usual gauge of gravitational waves. This is a consistency check since, as we saw in section \ref{sec:massless}, massless solutions (\ref{goodsols}) are general for the bilocal model.  

In summary, once we fix the appropriate gauge in gravity, the theories match perfectly, at least to linear order in perturbation theory.  

\subsection{Second and higher orders in perturbation theory}
\label{SHOPT}

In this subsection, we sketch how both theories, gravity and bilocal, should be compared at second and higher orders in the perturbative expansion. First, we review the perturbative expansion as usually performed in gravity.  

\subsubsection{Perturbative gravity}

In perturbation theory one assumes the existence of a one-parameter  family of solutions $g(\kappa)_{\mu\nu}$ to Einstein's equations
\begin{equation}\label{Einstein}
 R_{\mu\nu}-\frac{1}{2}g_{\mu\nu}R+\Lambda g_{\mu\nu}=8\pi G T_{\mu\nu}.
\end{equation}
The perturbation equations get generated as the Einstein's equations are expanded in powers of $\kappa$ and the terms with equal powers are equated. In this paper, for simplicity, we will consider pure gravity with no cosmological constant, so
\begin{equation}
\label{Einsteinflat}
G_{\mu\nu}\equiv  R_{\mu\nu}-\frac{1}{2}g_{\mu\nu}R =0.
\end{equation}
Now, both the solutions and the operator $G_{\mu\nu}$ are to be expanded in powers of $\kappa$. Any new solution of the $\kappa$-family\footnote{The parameter $\kappa$ is actually dimensionful. By consistency with the non-perturbative treatment it is found to be $\kappa=\sqrt{16\pi G}$, where $G$ is the Newton constant.} is expanded around flat metric as
\begin{equation}\label{solexp}
g_{\mu\nu}(\kappa)=\eta_{\mu\nu}+\kappa h^{(1)}_{\mu\nu}+\kappa^2  h^{(2)}_{\mu\nu}+\cdots ,
\end{equation}
where $h^{(i)}_{\mu\nu}$ are order $i$ contributions to the solution $g_{\mu\nu}(\kappa)$ and are found as solutions of some (linear) differential equations when solved in ascending order. To find out the tower of equations that $h^{(i)}_{\mu\nu}$  solves, we expand the Einstein's tensor. This is done by writing
\begin{equation}\label{opexp}
G_{\mu\nu}[\eta_{ab}+\kappa h_{ab}]=G^{(0)}_{\mu\nu}[\eta_{ab}]+\kappa G^{(1)}_{\mu\nu}[h_{ab}]+\kappa^2 G^{(2)}_{\mu\nu}[h_{ab}]+\cdots.
\end{equation}
A general order operator is them computed as
\begin{equation*}
 G^{(n)}_{\mu\nu}[h_{ab}]=\frac{1}{n!}\frac{d^n}{d\kappa^n}G_{\mu\nu}[\eta_{ab}+\kappa h_{ab}]\bigg|_{\kappa=0}.
\end{equation*}
Performing expansions (\ref{solexp}) and (\ref{opexp}) and equating equal powers of $\kappa$ we get the tower of equations
\begin{eqnarray}
0&=& G^{(0)}_{\mu\nu}[\eta_{ab}] \nonumber \\
0&=& G^{(1)}_{\mu\nu}[h^{(1)}_{ab}], \label{linear} \\
0&=& G^{(1)}_{\mu\nu}[h^{(2)}_{ab}]+ G^{(2)}_{\mu\nu}[h^{(1)}_{ab}],\label{secondorder}\\
&\vdots& \label{tower}
\end{eqnarray}
To find $h^{(2)}_{ab}$ one must solve (\ref{secondorder}) with the linear solutions $h^{(1)}_{ab}$ found by solving the first order.  $G^{(1)}$ is a linear differential operator. The operator $G^{(2)}[h]$ produces 24 terms which are schematically either of type $h\partial^2 h$ or type $(\partial h)^2$ with different index contractions. The usefulness of the perturbative method is that computing $h^{(1)}$ is just solving linear differential equations. Then, one computes $G^{(2)}[h^{(1)}]$ by inserting the obtained solutions. So  $G^{(2)}[h^{(1)}]$ is a known function and then equation (\ref{secondorder}) is just a set of linear differential equations for $h^{(2)}$. The solution will seed the third order perturbation equations, and so on. So, at end of the day, one can find approximate solutions to Einstein equations by solving, iteratively, sets of linear differential equations.

\subsubsection{Perturbative bilocal model}
As it is customary in perturbation theory, we expand the solutions in powers of $\kappa_B$, so
\begin{equation}\label{phiexp}
 \Phi=\Phi^{(1)}+\kappa_B\Phi^{(2)}+\kappa_B^2\Phi^{(3)}\dots
\end{equation}
An extension of the bilocal model should replace (\ref{eom}) by
\begin{equation}\label{genmodel}
\big(\Box+\Box_r-\alpha^4r^2+2\alpha^2\big)\Phi=V(\Phi),
\end{equation}
and (\ref{constraint}) by
\begin{equation}\label{genconstraint}
\frac{\partial}{\partial X^{\mu}}\bigg(\frac{\partial}{\partial r^{\mu}}+\alpha^2 r_{\mu}\bigg)\Phi(X,r)=W(\Phi).
\end{equation}
Potentials $V(\Phi)$ and $W(\Phi)$ will be expanded in powers of a coupling $\kappa_B$ with the correct dimensions\footnote{At first sight it could seem strange to plug a potential $W(\Phi)$ in (\ref{genconstraint}), since equation (\ref{genconstraint}) is a constraint. Actually, (\ref{genconstraint}) matches the Lorentz gauge condition for the free case. However, this constraint may change in gravity for higher orders of perturbation, as noticed by Wald \cite{W}. So the presence of a nonzero $W(\Phi)$  for higher orders in perturbation theory must be taken into account accordingly in the bilocal model.}. So, after plugging (\ref{phiexp}) into (\ref{genmodel}) and (\ref{genconstraint}) we get a tower of equations similar to (\ref{tower}), an equation for any power of $\kappa_B$. The claim we make in this paper is that after suitable choices of the potentials $V(\Phi)$ and $W(\Phi)$, solutions of gravity and bilocal should match order by order in $\kappa$, $\kappa_B$ being a function of $\kappa$. We have already checked it out for the first order, for which linearized gravity solutions, that is, solutions of (\ref{linear}), have been proven to match those of linear (\ref{genmodel}) and (\ref{genconstraint}), when their RHS is 0. We will now write
\begin{equation}\label{eomreloaded}
\big(\Box+\Box_r-\alpha^4r^2+2\alpha^2\big)\Phi^{(1)}(X,r)=0,
\end{equation}
and
\begin{equation}\label{constraintreloaded}
\frac{\partial}{\partial X^{\mu}}\bigg(\frac{\partial}{\partial r^{\mu}}+\alpha^2 r_{\mu}\bigg)\Phi^{(1)}(X,r)=0
\end{equation}
as first order equations.

The natural second order contribution for bilocal theories is a term like
\begin{equation}\label{int2}
\kappa_B\int \bar{\Phi}(x,z)\bar{\Phi}(z,y)dz,
\end{equation}
which is a function of $x$ and $y$, and so of $X$ and $r$.
The second order equation is obtained by equating terms proportional to $\kappa_B$ in (\ref{genmodel}) and (\ref{genconstraint}). For instance, from equation (\ref{genmodel}) we will have the $\kappa_B$-second order equation 
\begin{equation}\label{secondorderpert}
 \big(\Box+\Box_r-\alpha^4r^2+2\alpha^2\big)\Phi^{(2)}(X,r)=\int \bar{\Phi}^{(1)}(x,z)\bar{\Phi}^{(1)}(z,y)dz,
\end{equation}
where $\Phi^{(1)}$ are solutions of (\ref{eomreloaded}) and (\ref{constraintreloaded}). The functions $\Phi^{(2)}(X,r)$ lead, after derivation, to functions $H^{(2)}_{\mu\nu}(X)$ which will be identified with second order contributions in gravity $\bar{h}^{(2)}$, as we did in the linear case. 

The identification we claim is nontrivial. In gravity, the seed functions $h^{(1)}$ enter in the second order equations with their derivatives, whereas in the bilocal model they get integrated, as in (\ref{secondorderpert}). It is the special properties of Gaussians with respect to derivation and integration what are expected to make it possible.

\subsection{Physical interpretation of the bilocal model}
\label{PIBF} 

Let us speculate on the physical meaning of the bilocal field itself. 
First of all, if we demand that an observer cannot, {\it by any means}, resolve very small distances, a fundamental theory of gravity should take this limitation into account. 
This means that very close points should be identified from the macroscopic point of view, 
since they are indistinguishable. Of course, this restriction should vanish as we consider well-separated points. 
One may assign
\footnote{Note that this claim is compatible with the symmetry $\Phi(X,r)=\Phi(X,-r)$, assumed at the begining.} 
to a pair of points $(x,y)$ a probability $P(x=y)$ that decays with the distance between $x$ and $y$. The easiest probability distribution that makes the job with just one parameter, $\alpha$, is the Gaussian
\begin{equation*}
P(x=y)=e^{-\frac{\alpha^2}{2}d^2(x,y)},
\end{equation*}        
which decays exponentially for distances greater than $1/\alpha$. We speculate that 
such a probability is encoded in the bilocal field, once it is quantized. 
Along these lines, after quantization, we could also
interpret the bilocal field $\bar{\Phi}(x,y)$ 
as the probability amplitude of identifying points 
$x$ and $y$ when we try to resolve them.

On the other hand, the identification of spacetime points at small distances will result in a complex topology at those scales and, in turn, they should modify the usual commutator relations of matter fields. For instance, for a local scalar field $\phi$ one should replace
\begin{equation*}
[\phi(x),\phi^{\dagger}(y)]\propto \delta(x-y) \to [\phi(x),\phi^{\dagger}(y)]\propto P(x=y),
\end{equation*}
which ensures that for delta distributions we recover the usual commutator relations. 
This will apply to all matter fields as it should, since gravity should couple to everything.

\section{Conclusion}

In this paper we have investigated the relation between gravity and non-locality at the classical level. Non-locality has been  implemented by a minimal bilocal scalar field model and has been forced to be short-range. We have seen how bilocality introduces new (internal) degrees of freedom that can accommodate gravity. Specifically, we have shown that massless solutions of the free scalar field encode those of linearized gravity. To our knowledge, this is the first time that massless solutions of bilocal models have been studied. We have then shown how to proceed in order to match solutions in higher orders in perturbation theory, where there is self-interaction in both gravity and the bilocal field. 

We offer strong evidence for the emergence of gravity from non-locality. The claim is that full gravity can be obtained from interacting bilocal models as defined by equations (\ref{genmodel}) and (\ref{genconstraint}). If this is correct, then we believe that  the analysis of this paper goes beyond the bilocal field to general non-local fields. A non-local theory that is effectively local at large scales is expected to be effectively bilocal to leading order, much as the dipole term dominates a multipole expansion.

The classification of non-local theories can be made by considering the number of points that define the field. Therefore we have bilocal, trilocal..., and when the field depends on an infinite number of points, one can obtain 
classical string theory \cite{T2}. 
From the point of view of locality, bilocal theories are specially interesting since they encode departures from locality in the simplest form. We think that they deserve more attention.  In any case, it would be interesting to investigate  under which conditions a non-local theory leads to an effective bilocal theory. 

There are many lines along which this work can be extended.
From the classical point of view it would be interesting to test the match of gravity and bilocal field in second and higher orders \cite{WIP}. It is known that one can obtain a full theory of gravity in a unique way starting from linearized gravity and imposing reasonable consistency conditions \cite{D,W}. It would be interesting to know what these conditions translate into in the bilocal model. This, in the end, would tell us about the full shape of potentials $V(\Phi)$ and $W(\Phi)$ in equations (\ref{genmodel}) and (\ref{genconstraint}).

The background we have used as the starting point for the perturbative expansion is the Minkowski metric. It should not be hard to adapt the model to accommodate a cosmological constant, so that the expansions would be performed around de Sitter or anti-de Sitter backgrounds.

If solutions match it is natural to think that there should be some equivalence at the level of actions \cite{FP}. This should be useful for subsequent quantization. It would also shed light on the way general covariance enters on the bilocal field side. In the same spirit, gauge invariant variables could be considered.
Gauge invariant variables \cite{KI} have been proven to be extremely useful for computations in perturbation theory. It would be interesting to see how this formalism is realized in a bilocal model. It can potentially provide a  gauge invariant picture.

One of the initial motivations of this work was to find an appropriate setup for quantizing gravity. It is believed that the bilocal theories are, in general finite.
The absence of divergences is essentially due to the existence of a 
fundamental length. 
In this work, this length scale is $\alpha$, which measures non-locality. Once it is understood how bilocal fields encode gravity, it seems possible to keep track of the ``induced'' metric field after quantizing the bilocal field. We hope that
this points the way toward an eventual finite theory of quantum gravity. 

\vspace{0.2cm}
\noindent
{\bf Acknowledgment}

\noindent
We thank J.L. Cortes, G. Kunstatter, F. Mir and A. Segui for useful discussions. 
This work is supported by the 
Natural Sciences and Engineering Research Council of Canada and the 
University of Lethbridge.

\end{document}